# Dynamic Imprints of Colliding-wind Dust Formation from WR 140

Emma P. Lieb[1], Ryan M. Lau[2], Jennifer L. Hoffman[1], Michael F. Corcoran[3,4], Macarena Garcia Marin[5],
Theodore R. Gull[4], Kenji Hamaguchi[4,6], Yinuo Han[7], Matthew J. Hankins[8], Olivia C. Jones[9], Thomas I. Madura[10],
Sergey V. Marchenko[11], Hideo Matsuhara[12], Florentin Millour[13], Anthony F. J. Moffat[14], Mark R. Morris[15], Patrick W. Morris[7],
Takashi Onaka[16], Marshall D. Perrin[17], Armin Rest[17], Noel Richardson[18], Christopher M. P. Russell[19],
Joel Sanchez-Bermudez[20], Anthony Soulain[21], Peter Tuthill[21], Gerd Weigelt[22], and Peredur M. Williams[23]

[1] University of Denver, Denver, CO 80210, USA; emma.lieb@du.edu
[2] NOIRLab, Tucson, AZ 85719, USA
[3] Catholic University of America, Washington, DC 20064, USA
[4] NASA Goddard Space Flight Center, Greenbelt, MD 20771, USA
[5] European Space Agency, Space Telescope Science Institute, Baltimore, MD 21218, USA
[6] University of Maryland Baltimore County, Catonsville, MD 21250, USA
[7] California Institute of Technology, Pasadena, CA 91125, USA
[8] Arkansas Tech University, Russellville, AR 72801-2222, USA
[9] United Kingdom Astronomy Technology Centre, Edinburgh, EH9 3HJ, UK
[10] San Jose State University, San Jose, CA 95112, USA
[11] Science Systems and Applications, Inc., Lanham, MD 20706, USA
[12] Institute of Space & Astronautical Science, Kanagawa, 252-5210, Japan
[13] Observatoire de la Cote d'Azur, Nice, 06300, France
[14] Universite de Montreal, Montreal, Quebec, H2V 0B3, Canada
[15] University of California—Los Angeles, Los Angeles, CA 90095, USA
[16] Department of Astronomy, Graduate School of Science, University of Tokyo, Tokyo 181-0015, Japan
[17] Space Telescope Science Institute, Baltimore, MD 21218, USA
[18] Embry-Riddle Aeronautical University, Prescott, AZ 86301, USA
[19] Department of Physics and Astronomy, Bartol Research Institute, University of Delaware, Newark, DE 19716, USA
[20] Instituto de Astronomia, Universidad Nacional Autonoma de Mexico, 04510 Ciudad de México, CDMX, Mexico
[21] University of Sydney, Sydney, New South Wales, NSW 2006, Australia
[22] Max Planck Institute for Radio Astronomy, Bonn, D-53121, Germany
[23] Institute for Astronomy, University of Edinburgh, Royal Observatory, Edinburgh, EH9 3HJ, UK
Received 2024 October 18; revised 2024 November 22; accepted 2024 November 25; published 2025 January 13

## Abstract

Carbon-rich Wolf–Rayet (WR) binaries are a prominent source of carbonaceous dust that contribute to the dust budget of galaxies. The "textbook" example of an episodic dust-producing WR binary, WR 140 (HD 193793), provides us with an ideal laboratory for investigating the dust physics and kinematics in an extreme environment. This study is among the first to utilize two separate JWST observations, from Cycle 1 ERS (2022 July) and Cycle 2 (2023 September), to measure WR 140's dust kinematics and confirm its morphology. To measure the proper motions and projected velocities of the dust shells, we performed a novel point-spread function (PSF) subtraction to reduce the effects of the bright diffraction spikes and carefully aligned the Cycle 2 to the Cycle 1 images. At 7.7 μm, through the bright feature common to 16 dust shells (C1), we find an average dust shell proper motion of $390 \pm 29\,\mathrm{mas\,yr^{-1}}$, which equates to a projected velocity of $2714 \pm 188\,\mathrm{km\,s^{-1}}$ at a distance of 1.64 kpc. Our measured speeds are constant across all visible shells and consistent with previously reported dust expansion velocities. Our observations not only prove that these dusty shells are astrophysical (i.e., not associated with any PSF artifact) and originate from WR 140, but also confirm the "clumpy" morphology of the dust shells, in which identifiable substructures within certain shells persist for at least 14 months from one cycle to the next. These results support the hypothesis that clumping in the wind collision region is required for dust production in WR binaries.

*Unified Astronomy Thesaurus concepts:* Wolf-Rayet stars (1806); Binary stars (154); Dust formation (2269)

*Materials only available in the* online version of record: animation

## 1. Introduction

Wolf–Rayet (WR) binary stars of the carbon-rich subtype (WC) are believed to be significant producers of carbonaceous dust. Dust is created in WC binaries where the hot, dense, hydrogen-poor stellar wind of the WC star collides with the weaker hydrogen-rich wind of the OB companion, creating a shock cone in which the conditions are ideal for the condensation of the carbon-rich stellar material into dust particles (V. V. Usov 1991; S. V. Marchenko et al. 2003). Systems in which we can resolve the dusty nebula surrounding the central inner colliding-wind region provide us with astrophysical laboratories for understanding the production physics and propagation kinematics of dust from WC binaries (P. M. Williams & P. R. J. Eenens 1989; P. M. Williams et al. 2009). A quintessential example of a periodic, episodic WC dust-producing binary is the system WR 140 (HD 193793).

WR 140 consists of an evolved WC7 star and an early-type O5.5fc companion in a highly eccentric orbit ($e = 0.8993 \pm 0.0013$; J. D. Thomas et al. 2021). These massive stars orbit each other in 2895 days (7.93 yr). An optimal distance of $1.64^{+0.08}_{-0.07}$ kpc, which we adopt here, was calculated by C. A. L. Bailer-Jones et al. (2018) from Gaia DR2 with Bayesian inference (J. D. Thomas et al. 2021). Unlike WC binaries that produce dust continually (P. G. Tuthill et al. 1999), WR 140 produces most of its dust over the course of a few months when the stars approach periastron passage, which lasts more than half the angular orbit due to the orbit's high eccentricity. Dust production only occurs near periastron because it requires the hot dense wind of the WR star to be sufficiently compressed by that of the O star companion, yet most of the dust that accumulates around the stars appears somewhat later in the orbit, because dust is spewed out more slowly after periastron (Y. Han et al. 2022). JWST Cycle 1 imaging (R. M. Lau et al. 2022) revealed 17 individual shells that have persisted for over ~130 yr. The dusty, IR-emitting nebula created from this type of episodic dust production resembles segmented dust shells, which may propagate at varying projected velocities due to changes in the properties of the interacting winds as well as their angle of propagation relative to our line of sight (R. M. Lau et al. 2023). Y. Han et al. (2022) showed that the creation of these shells must begin with an initial burst of dust grains produced near periastron, which is compressed into shells by the strong colliding winds. These studies provided valuable information about the structure and formation of the shells. Now, by analyzing a second set of JWST images, we investigate how the physics driving dust production affects the large-scale kinematics of the resulting dust structures.

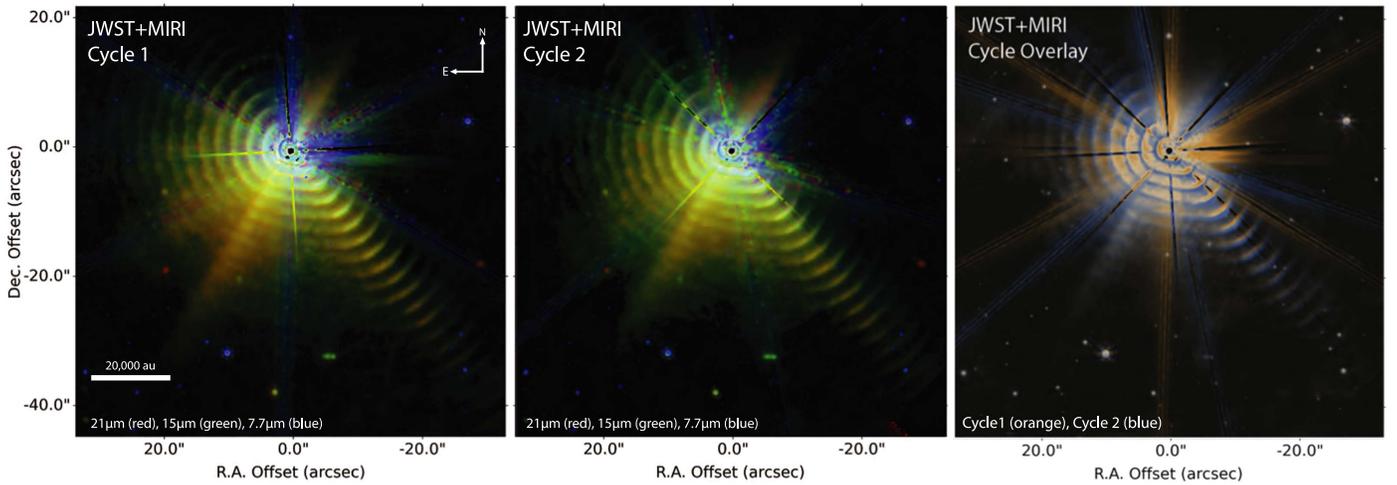

**Figure 1.** False-color red, green, blue and orange–blue overlay. Point-spread function (PSF) subtracted Cycle 1 image (left), Cycle 2 image (middle), and comparison image of Cycle 1 (orange) to Cycle 2 (blue) at 7.7 $\mu$m (Section 3.2). The alignment of the two cycles to each other is highlighted in the background stars of the right panel (cycle overlay); their white color indicates that the orange Cycle 1 image is well aligned with the blue Cycle 2 image (Section 3.3). See the Appendix for GIFs that blink between the two cycles, illustrating the motion of the shells.

Previous ground-based observations (J. D. Monnier et al. 2002; S. V. Marchenko & A. F. J. Moffat 2007; P. M. Williams et al. 2009; P. Williams 2011) resolved one to two of the bright shells near the star, enough to show them forming periodically near periastron, but higher-resolution, large-field, low-background imaging from space with JWST was necessary to characterize the physical extent of the fainter outer shells. These images reveal the radial proper motion of the series of dust shells around a prototypical WR binary as well as highlighting the clumpy morphology of the dusty shells. Bright sources, like WR 140, oversaturate the Mid-Infrared Instrument (MIRI) detector and cause diffraction spikes that interfere with our science goals. To mitigate the effects of these bright artifacts, we performed a subtraction of the diffraction spikes in all images using a MIRI detector-sampled model of the point-spread function (PSF). Additionally, we rigorously aligned the Cycle 2 images to the Cycle 1 images such that any change between them is due only to the movement of the dust. With those steps accomplished, we here compare the original JWST images from R. M. Lau et al. (2022) to our new observations taken a year later to examine the motion and time-dependent morphology of WR 140's dusty architecture (Figure 1).

This Letter is structured as follows. In Section 2 we describe the observing parameters used to capture the images. In Section 3 we discuss how we reduced and analyzed our images. In Section 4 we compare the Cycle 1 and Cycle 2 images quantitatively to investigate the dynamics and time-dependent morphology of the dust shells. Last, in Section 5, we summarize this work, discuss the implications of our results, and outline the valuable avenues of research still to be explored.

## 2. Observations

The two sets of observations we present in this work were taken with JWST's Mid-Infrared Instrument (MIRI; G. S. Wright et al. 2015). The filters used for both sets of images were F770W, F1500W, and F2100W, which correspond to wavelengths of 7.7, 15, and 21 $\mu$m, respectively; these IR filters probe the thermal dust emission around WR 140. The first set of images was taken on 2022 July 27 during the Director's Discretionary Early Release Science program (ID = ERS 1349, hereafter referred to as Cycle 1) and corresponds to an orbital phase of 0.7, roughly 5.5 yr after the last periastron passage. The second was taken on 2023 September 8 during Cycle 2 and corresponds to an orbital phase of 0.9, roughly 6.7 yr after the last periastron passage (ID = 3823). Each set of observations had a total exposure time of ~24 minutes (1431.924 s) and employed three sets of four-point dithers to improve PSF sampling, reduce detector artifacts, and enhance contrast.

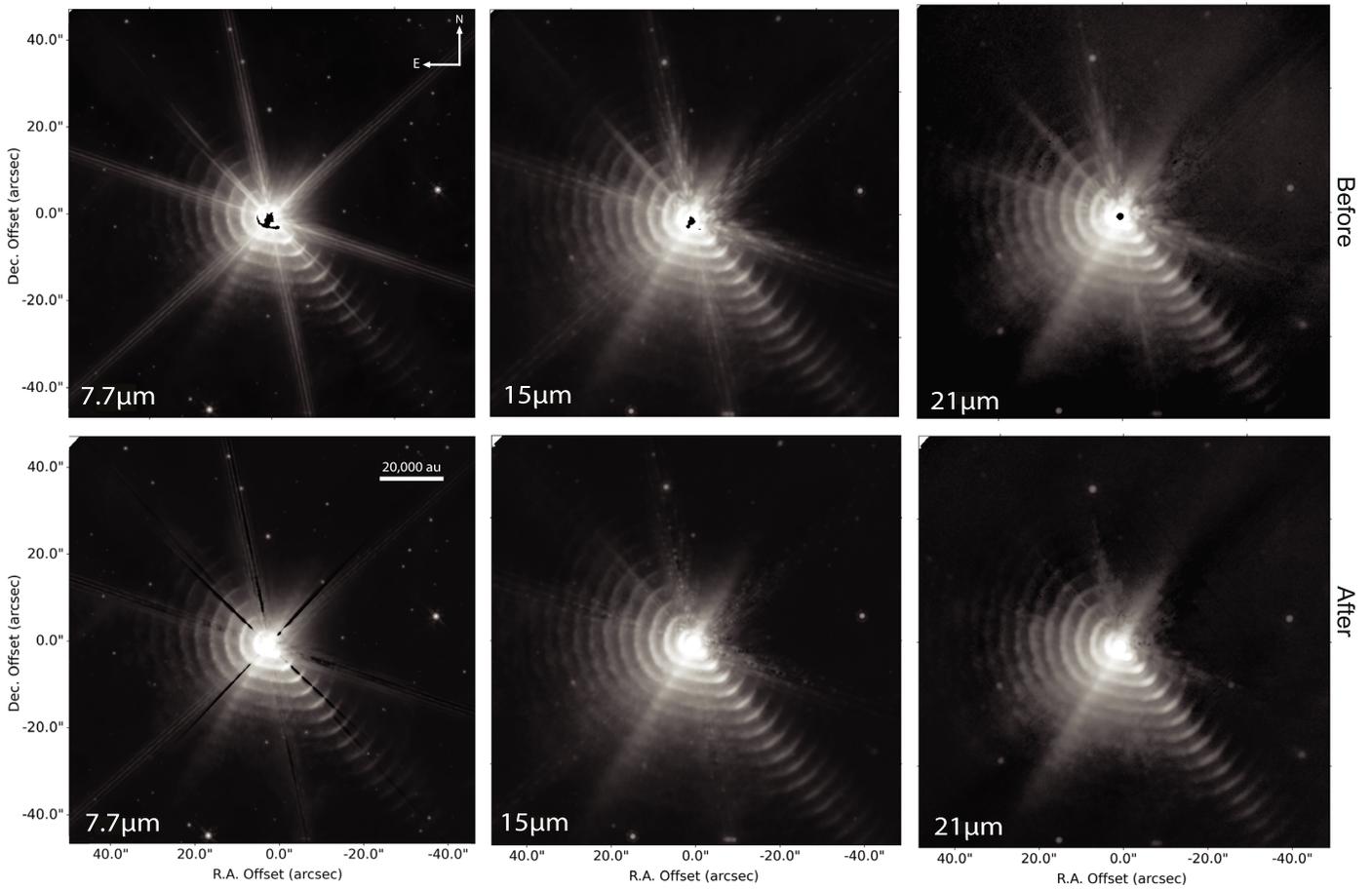

**Figure 2.** PSF subtraction. Before (top) and after (bottom) PSF subtraction of Cycle 2 images at 7.7, 15, and 21 $\mu$m (Section 3.2). The varying flux of the diffraction spikes as a function of radius in the 7.7 $\mu$m image required a complex scaling of the PSF model, which resulted in some areas of oversubtraction. See the Appendix for the corresponding figure for the Cycle 1 images.

## 3. Methods

### 3.1. Data Reduction

Due to the bright central core of the binary, the innermost shells are saturated in our image reductions using the default pipeline parameters. We were able to reduce some of the effects of this saturation by rerunning our own Stage 1 and Stage 2 reductions from the uncalibrated data, using pipeline version 1.16.0 and CRDS context 1293. We changed the calibration parameters from the default pipeline to reduce the growth of saturated pixels. Specifically, we skipped the first frame step, set the parameter suppress_one_group=False for the ramp_fit step, and set n_pix_grow_sat=0 for the jump step. Collectively, these adjustments enabled the pipeline to make flux estimates even from pixels that saturated after only a single read at the start of the ramp, which was the case for the very bright inner regions around WR 140. Such estimates are inherently less precise than ramp fits to less extremely illuminated pixels, but we found empirically for the WR 140 data that the inner pixel measurements obtained in this way are sufficiently precise and are preferable to leaving those pixels as missing values. This step was crucial to finding a precise centroid for our target in each dither frame so that the center of the PSF model we created could be matched optimally to the intersection point of the diffraction spikes through the source.

### 3.2. PSF and Background Subtraction

We then performed the PSF subtraction on our newly created Stage 2 data. In order to analyze the details of WR 140's dust structures in our observations, it was necessary to reduce the effects of the bright diffraction spikes (Figure 2, top). The shape of the MIRI PSF arises from multiple effects: diffraction from the hexagonal primary mirror causes six large diffraction spikes, while two smaller spikes are caused by the secondary mirror support. At wavelengths below 10 $\mu$m, an additional four smaller perpendicular spikes (denoted as the "cruciform") arise from internal scattering within the MIRI detectors (A. Gáspár et al. 2020; D. Dicken et al. 2024). We used WebbPSF (M. D. Perrin et al. 2012) to create our PSF model, which was normalized to unity and therefore needed to be scaled in flux to match the science target.

We first subtracted the background to accurately scale up the PSF model to the source data flux, since the background flux at the position of the spikes affects the PSF scaling. To do this, we performed an iterative background estimation using the background library from the photutils Python package. First, we estimated the background of the science image using a box size of 15 × 15 pixels (see "Background2D" from photutils for more details; L. Bradley et al. 2024). From this estimate, we calculated a mask such that fluxes above a threshold were set to false. We did this to ensure that at the inner regions of the dust structure we did not subtract away real dust emission, but only the background. Once we created this mask, we calculated a new background using a box size of 50 × 50 pixels, with a fill value equal to 0.01 so that the edges of the mask were similar in flux to the background there, preventing a sharp boundary. Then we subtracted this background from the science image and moved on to scaling up the PSF model.

To find a systematic scale factor for the shortest wavelength filter, 7.7 μm, we calculated the flux of the image along a circular path defined by the centroid of our target and an arbitrarily large radius, such that the flux peaks along this path corresponded to the diffraction spikes crossing through the circle. From those pixel coordinates, we calculated the ratio of the science image fluxes at the spikes to the PSF model fluxes at the spikes. We then used the average of the ratios for all the spike pixels to scale up the PSF model to match the flux level of the science image. For the two longer-wavelength filters, 15 and 21 μm, we found a sufficient scale factor by taking a fraction (1/5 and 1/8, respectively) of the ratio of the brightest flux in the science image to the brightest flux in the PSF model. After subtracting the PSF from each dither frame for each filter for both cycles, we then aligned the Cycle 2 images to the World Coordinate System (WCS) of the Cycle 1 images.

### 3.3. Image Alignment

JWST MIRI images can be aligned to a WCS with the package JHAT (A. Rest et al. 2023a, 2023b). JHAT performs a relative alignment by taking the first image (Cycle 1 in our case) and aligning the second image (Cycle 2) to it. It does so by locating background point sources in the first image and creating a catalog of those stars to search for in the second image. Our final images are aligned to within 0.1 pixels, which corresponds to a maximum distance uncertainty of 11 mas or 18 au.[24]

In our data, the 15 and 21 μm images have substantially fewer field stars than the 7.7 μm image, which led to unsatisfactory initial alignments. To account for this, we copied the new JHAT-aligned WCS of the 7.7 μm image to the 15 and 21 μm images. This ensured that all three wavelengths referred to the same WCS and that the registration from Cycle 1 to Cycle 2 was consistent across all images. Finally, we ran the JWST pipeline to create Level 3 mosaics (Figure 2, bottom), which we used for the measurements we discuss in the following sections.

## 4. Results and Discussion

The nested shells of carbon-rich dust surrounding WR 140 in the Cycle 2 images have a very similar morphology to those in our Cycle 1 images, taken 1.12 yr earlier (Figure 1). We see the same arrangement of concentric shells with an average separation of 2″.67 (4380 au at our assumed distance of 1.64 kpc; G. Rate & P. A. Crowther 2020; J. D. Thomas et al. 2021). However, these shells have moved radially outward since the Cycle 1 observations (Figure 1). Their persistent morphologies in Cycle 2 observations, 14 months after their confirmation by R. M. Lau et al. (2022), prove that they are astrophysical structures associated with dust production from the binary, and not a result of any artifacts resulting from the PSF diffraction spikes. We focus on three dust structures denoted as "C1," "C0," and "E" (P. M. Williams et al. 2009; Figure 3). The direction of the C1 feature hosts the most visible shells and is oriented to the southwest of WR 140. The C0 feature is seen as a linear spur-like structure located toward the southeast of WR 140. E is named for the "Eastern arm" and is located to the east of WR 140. These three dust features are visible in each shell, and each moves through space along a different trajectory, as shown by the geometrical modeling of Y. Han et al. (2022) and R. M. Lau et al. (2022). The C1 and C0 features move predominantly in the plane of the sky, so that their projected velocities are approximately the same as their full space velocities, and any radial velocity component is negligible. The E feature, on the other hand, has a considerable velocity component along our line of sight, which means we must deproject the measured image velocities of this feature to convert them to full space velocities (Section 4.1). In this process, we also convert from pixel space to physical space, which depends on the distance to WR 140. We chose to use the 1.64 kpc distance calculated by C. A. L. Bailer-Jones et al. (2018) rather than one derived from the more recent Gaia DR3 (Gaia Collaboration et al. 2023) to keep our analysis consistent with that of R. M. Lau et al. (2022) and to avoid any potential errors in that data set. We report below the projected space velocities of C1 and C0 and the deprojected space velocities of E.

### 4.1. Proper Motions

To measure the proper motions of the dust shells between Cycles 1 and 2, in each of our images we defined ten 3 pixel wide rays from the position of WR 140, spread out radially toward the southwest direction (∼220°) along the C1 feature (Figure 3) out to a radius of ∼40″, which pass through ∼16 shells each. We then took the median flux density values across all 10 rays and plotted the resulting radial flux profiles to quantify the locations of the shells (Figure 4). This method of using several rays increases signal to noise as well as reducing the effects from background sources and artifacts from the PSF subtraction.

Using these data, we then calculated the distance each shell moved over the course of the 1.12 yr between the two observations. We first divided out the slope visible in Figure 4 to normalize the profile for each wavelength and dust feature before measuring the peak positions. To define the slope, we utilized the fit_continuum function from the specutils Python package (N. Earl et al. 2024) with its default parameters of a Chebyshev1D model and a trust-region reflective algorithm and least-squares statistic (TRFLSQFitter) fitter. After normalization, we then fitted each peak in each radial flux profile with a Gaussian to measure the precise position of each shell in both cycles. Finally, we subtracted the Cycle 1 position from the Cycle 2 position, converted the differences from pixels to mas and km, and divided by the time difference between observations to obtain the proper motions in mas yr$^{-1}$ and the projected velocities in km s$^{-1}$. We list the cycle-to-cycle separations, proper motions, and projected/deprojected velocities in Table 1 in the Appendix. From this analysis, we find that the median proper motions of the shells in C1 are 332 ± 20 mas yr$^{-1}$ at 7.7 μm, 302 ± 14 mas yr$^{-1}$ at 15 μm, and 327 ± 8 mas yr$^{-1}$ at 21 μm. At a distance of 1.64 kpc, these motions correspond to median projected velocities of 2586 ± 152 km s$^{-1}$ at 7.7 μm, 2350 ± 109 km s$^{-1}$ at 15 μm, and 2547 ± 61 km s$^{-1}$ at 21 μm (Figure 5, top). These speeds are only slightly less than the estimated terminal wind speed of the WC7 star in WR 140 (2860 km s$^{-1}$; P. R. J. Eenens & P. M. Williams 1994) and are generally consistent with the dust stream velocity calculated by R. Fahed et al. (2011) at 2170 ± 100 km s$^{-1}$. We note that an increase in the assumed WR 140 distance of 10% (as potentially implied by the Gaia DR3 parallax measurements) would result in an increase in our calculated projected velocities of ∼200 km s$^{-1}$. This would bring them into closer alignment with the terminal wind speed (P. R. J. Eenens & P. M. Williams 1994).

---
[24] The conversion from pixels to km or au is distance dependent; we adopt a distance of 1.64 kpc for WR 140 (J. D. Thomas et al. 2021; Section 4).

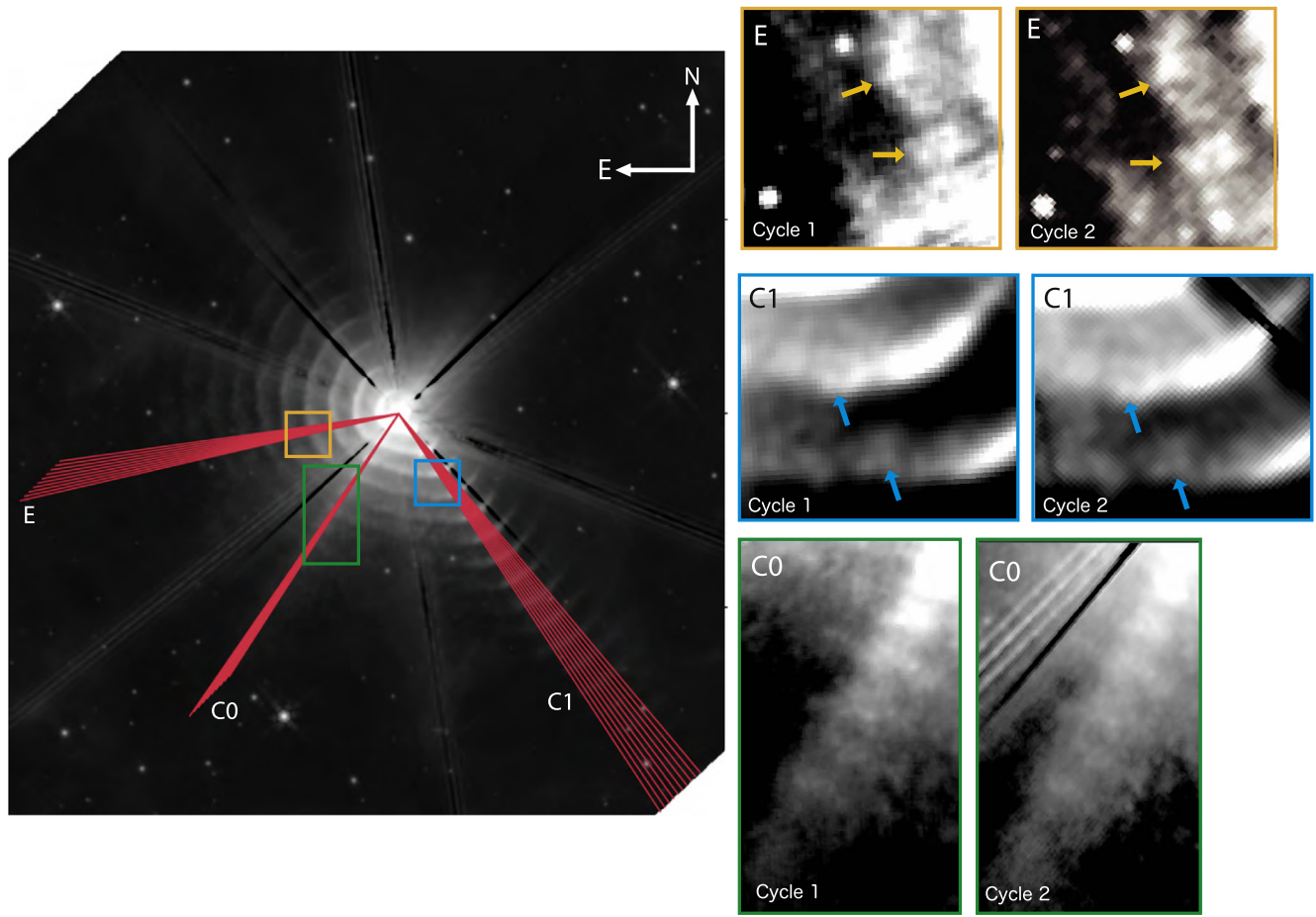

**Figure 3.** Substructure persisting in WR 140. Left: 7.7 μm PSF-subtracted Cycle 2 image with red lines indicating the 10 radial flux profiles per dust structure (E, C0, and C1) that we used to calculate the proper motions (Section 4.1). The three colored boxes correspond to the zoomed-in regions shown on the right (yellow for E, blue for C1, and green for C0). Top right: zoomed-in 7.7 μm images for Cycle 1 (left) and Cycle 2 (right) at the E dust structure, with clumps highlighted by arrows. Middle right: as above, but for the C1 dust structure. Bottom right: as above, but for the C0 dust structure. The approximate position angles of these dust features are 220°, 150°, and 115° east of north for C1, C0, and E, respectively.

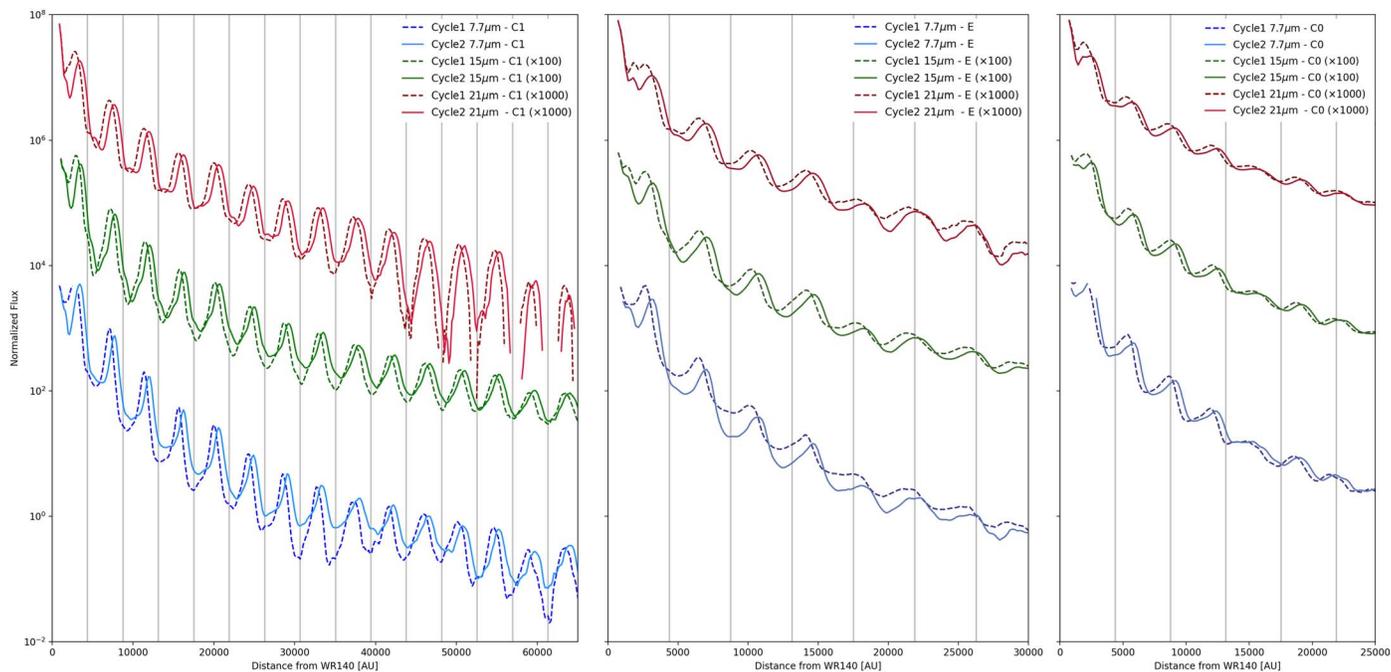

**Figure 4.** Radial flux profiles. Median of 10 radial flux profiles for Cycles 1 and 2 at 7.7, 15, and 21 μm through the major dust structures: C1, E, and C0 from left to right (Section 4). Gray vertical lines indicate intervals of 2″.67 (4380 au at our assumed distance of 1.64 kpc), which is the median separation between the flux density peaks of C1 (R. M. Lau et al. 2022).

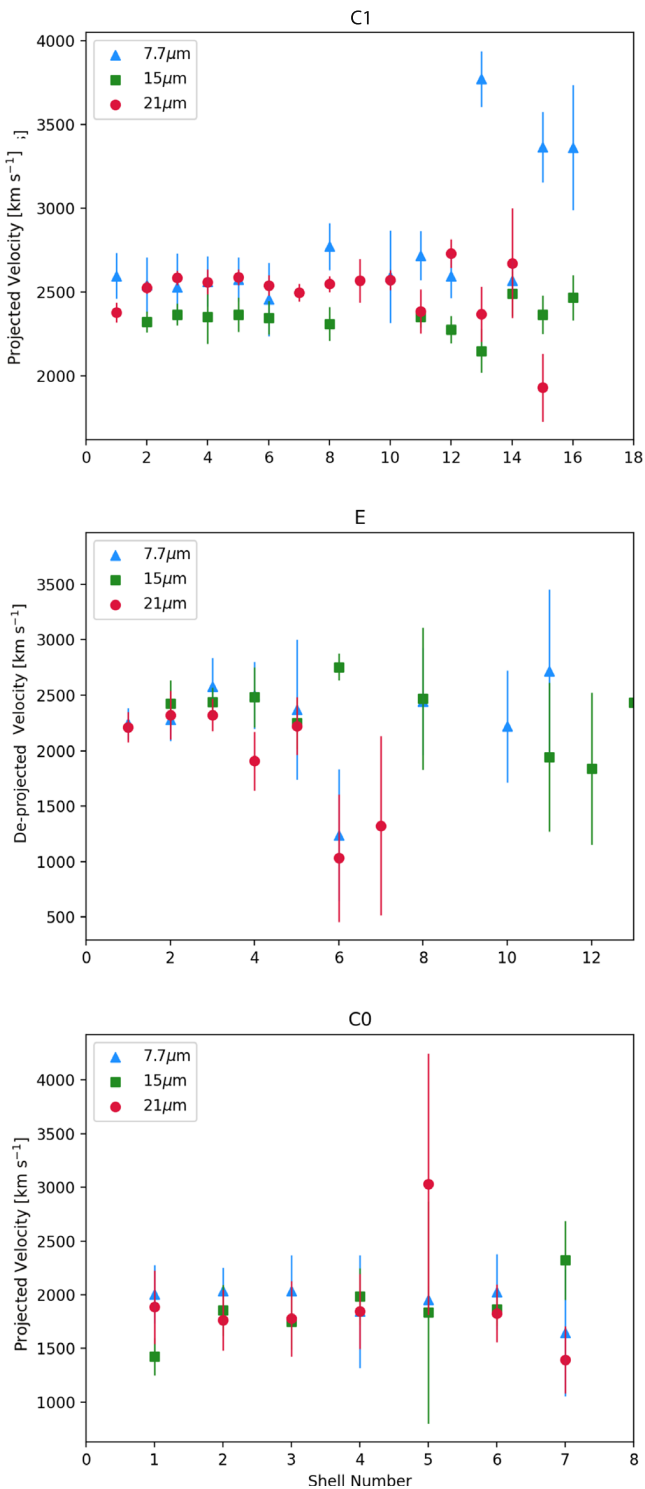

**Figure 5.** Speed vs. shell. Linear speeds measured through the C1 feature (top), the E feature (middle; deprojected), and the C0 feature (bottom). See Section 4.1 for details of the calculation and Table 1 in the Appendix for values and uncertainties.

In addition to C1, we also calculated radial flux profiles for ten 3 pixel wide rays starting at the position of WR 140, toward the eastward direction (∼115°) along the E feature out to a radius of ∼20″. Along this path, the profile passes through ∼11 visible shells. This region of the nebula has a significant velocity component toward the observer (Y. Han et al. 2022; R. M. Lau et al. 2022, 2023), so the speeds we present for E have been deprojected to only report the true space velocity. The deprojected velocity, $v'$, is found from

$$v' = v(\sec(180° - i - \theta)), \qquad (1)$$

where $v$ is the median projected velocity from the radial flux profiles, $i$ is the inclination angle (119°.6 ± 0°.5; R. Fahed et al. 2011), and $\theta$ is the half-opening angle of the wind shock cone (40° ± 5°; Y. Han et al. 2022). Along E, we find median proper motions of 289 ± 63 mas yr$^{-1}$ at 7.7 μm, 302 ± 35 mas yr$^{-1}$ at 15 μm, and 271 ± 31 mas yr$^{-1}$ at 21 μm, which correspond to median deprojected velocities of 2252 ± 490 km s$^{-1}$ at 7.7 μm, 2347 ± 275 km s$^{-1}$ at 15 μm, and 2107 ± 241 km s$^{-1}$ at 21 μm (Figure 5, middle). Finally, we repeated the analysis with 10 rays through the C0 feature, toward the southeast (∼150°). The shells are less visible along this feature, but we detected ∼7 peaks in the radial flux profile out to a radius of ∼15″. The median proper motions along C0 are 258 ± 43 mas yr$^{-1}$ at 7.7 μm, 239 ± 33 mas yr$^{-1}$ at 15 μm, and 235 ± 44 mas yr$^{-1}$ at 21 μm, which correspond to median projected velocities of 2005 ± 334 km s$^{-1}$ at 7.7 μm, 1856 ± 259 km s$^{-1}$ at 15 μm, and 1826 ± 340 km s$^{-1}$ at 21 μm (Figure 5, bottom). We found no statistically significant dependence of projected (or deprojected) velocity on wavelength; the speeds for the three wavelengths are consistent within standard uncertainties.

Figure 5 shows that at each dust feature and for each wavelength, our calculated expansion velocities are broadly consistent across all shells. Outlying measurements exhibit large uncertainties, likely due to low signal to noise in the outermost shells as well as the effect of background stars on the radial flux profiles. The consistency of the speeds as a function of radius implies that there is little effect on the expansion rate of the dust shells from the local interstellar medium (ISM); thus, the steady outflow of stellar winds after the dust formation period is the primary driver of the dust shell expansion.

Our results show a clear discrepancy in speed between the three dust features, in agreement with P. M. Williams et al. (2009, their Table 5), who showed that the three persistent dust features vary in proper motion, with C0 the slowest and C1 the fastest. This discrepancy among the three features may be due to a variation in dust grain size throughout the dust formation period of WR 140's periastron passage, in addition to different ratios of radial and tangential velocities. The speeds we measure in the shells at radii exceeding 10″ represent the terminal drift velocity of the dust grains, which is reached when the radiation pressure from the stellar winds is in equilibrium with the drag force from the ambient gas already present in the nebula. Both the force on the dust grains from the radiation pressure and the drag force are proportional to the dust grains' projected area (B. T. Draine 2011). The result that C0 exhibits a significantly slower proper motion than the other two dust features may indicate that the dust grains in that region are smaller, and therefore formed later than those in E (P. M. Williams et al. 2009). R. M. Lau et al. (2023) also showed through dust modeling that the grains in C0 are likely smaller than those in C1, which further supports the conclusion that the velocity difference between the three features is real. In a follow-up paper, we plan to present a detailed analysis of the dust grain properties around WR 140, including measuring dust density as a function of radius and characterizing the accelerations and decelerations of the dust grains. The slower speed of C0 could also be due in part to a projection effect; this possibility requires further investigation.

### 4.2. Confirmation of Nonuniform Shell Substructure

The Cycle 2 images also show that the dust shells produced by WR 140 are not perfectly uniform. Each shell in our images exhibits some degree of nonuniform substructure, and this substructure persists for at least 1.12 yr, from one set of images to the next (Figure 3). From these images, we can place an upper limit on the clump size of the MIRI PSF FWHM at 7.7 $\mu$m, which is $0\farcs269$ or 441 au at a distance of 1.64 kpc. Their endurance over the time between observations suggests that the clumpy substructures around WR 140 are indicators of the underlying dust formation mechanism.

Clumping due to compressible turbulence can occur in both stellar winds, though the turbulence in the $\sim 10\times$ stronger wind of the WR star will dominate in the formation of clumps (S. Lépine et al. 1999). Irregularities in the dust structure can also arise in the wind collision region, which tends to first obliterate the incoming wind clumps (J. Pittard 2006; J. M. Pittard 2007). J. W. Eatson et al. (2022) showed through numerical simulation of WR 140 around periastron that there is a significant relationship between the dust production rate and the orbital separation. As the two stars move in toward one another during periastron, the shocked wind-collision region changes from initially adiabatic to a radiative, high-density state with instabilities ideal for dust production. The instabilities in the wind-collision region produce bursts or clumps of dust rather than a continuous stream of dust grains throughout the period of dust production. Our confirmation of nonuniform substructure within the dust shells far from the central binary supports the hypothesis that clumps are needed to enable dust formation in the colliding winds.

### 5. Conclusion

We compared a second set of JWST MIRI images of WR 140 (Cycle 2) with those taken in Cycle 1, 14 months earlier (R. M. Lau et al. 2022). Our image analysis consisted of the following:

1. Subtracting the bright PSF diffraction spikes using a WebbPSF detector-sampled model;
2. Aligning the Cycle 2 images to the Cycle 1 images to within 0.1 pixels using JHAT and applying the optimally aligned WCS to all images;
3. Measuring the proper motions and projected velocities of the shells in the regions corresponding to the most notable dust features (C1, C0, and E); and
4. Comparing the detailed morphologies of the brightest shells between the two cycles.

We showed that the dust shells of WR 140 are expanding at steady velocities consistent with that of its WC7 stellar wind. We find little evidence of acceleration or deceleration of the dust shells. The varying speeds through the different morphological features of WR 140 may hold information about the nature of the dust in those regions. Our observations also confirm that the clumpy morphology of each shell survives for at least a year's time and suggest that clumps, which may arise from instabilities in the wind-collision region, are an important factor for dust formation.

To characterize accelerations or decelerations in all parts of the dust structure, not limited to the three features we explored here, it would be valuable to perform a similar analysis to ours using rays in a full circle sweeping around WR 140. Additionally, a worthwhile observation to explore in future work would be higher spatial resolution imaging to accurately determine the physical size of the clumpy substructure. Finally, characterizing the full extent of the bubble nebula that houses WR 140 would allow us to establish the timescale on which its dust enriches the ISM.

Our results highlight that the mechanisms for dust formation in WC binaries leave behind traces even in the farthest reaches of the dusty architecture of a system like WR 140. The steady expansion of dust away from WR 140 suggests that the immediate stellar environment has a low density; WR 140 may sit inside a cavity cleared out by its own WR wind.

### Acknowledgments

The work of E.P.L. is supported by NOIRLab, which is managed by the Association of Universities for Research in Astronomy (AURA) under a cooperative agreement with the U.S. National Science Foundation. J.L.H. acknowledges support from the National Science Foundation under award AST-1816944. T.O. acknowledges support by the Japan Society for the Promotion of Science (JSPS) KAKENHI grant No. JP24K07087. N.D.R. is grateful for support from the Cottrell Scholar Award #CS-CSA-2023-143 sponsored by the Research Corporation for Science Advancement. J.S.-B. acknowledges the support received from the UNAM PAPIIT project IA 105023. C.M.P.R. acknowledges support from NASA Chandra Theory grant TM3-24001X. This material is based upon work supported by NASA under award number 80GSFC24M0006 and based on observations made with the NASA/ESA/CSA James Webb Space Telescope. The data were obtained from the Mikulski Archive for Space Telescopes at the Space Telescope Science Institute, which is operated by the Association of Universities for Research in Astronomy, Inc., under NASA contract number NAS 5-03127 for JWST. These observations are associated with programs #3823 and #1349. Support for program #3823 was provided by NASA through a grant from the Space Telescope Science Institute, which is operated by the Association of Universities for Research in Astronomy, Inc., under NASA contract NAS 5-03127. We thank the anonymous reviewer for insightful feedback that improved the quality of this manuscript. We thank Christopher Packham and Mason Leist for their valuable discussions about the MIRI PSF subtraction.

### Data and Code Availability

This research made use of Astropy, a community-developed Python core package (Astropy Collaboration et al. 2022). The data used in this Letter are publicly available on the Barbara A. Mikulski Archive for Space Telescopes (MAST) portal. The observations are associated with proposal IDs 1349 for Cycle 1 and 3823 for Cycle 2, all data can be found at doi:10.17909/z8rb-ns85. The code used to perform the PSF subtraction can be found at this GitHub repository[25] (E. Lieb 2024).

## Appendix

This Appendix contains the before and after comparison of the PSF and background subtraction image processing for the Cycle 1 images (Figure 6), a table of data that is visualized in Figures 4 and 5 (Table 1), and two videos that blink between each cycle showing the motion of the dust shells (Figure 7).

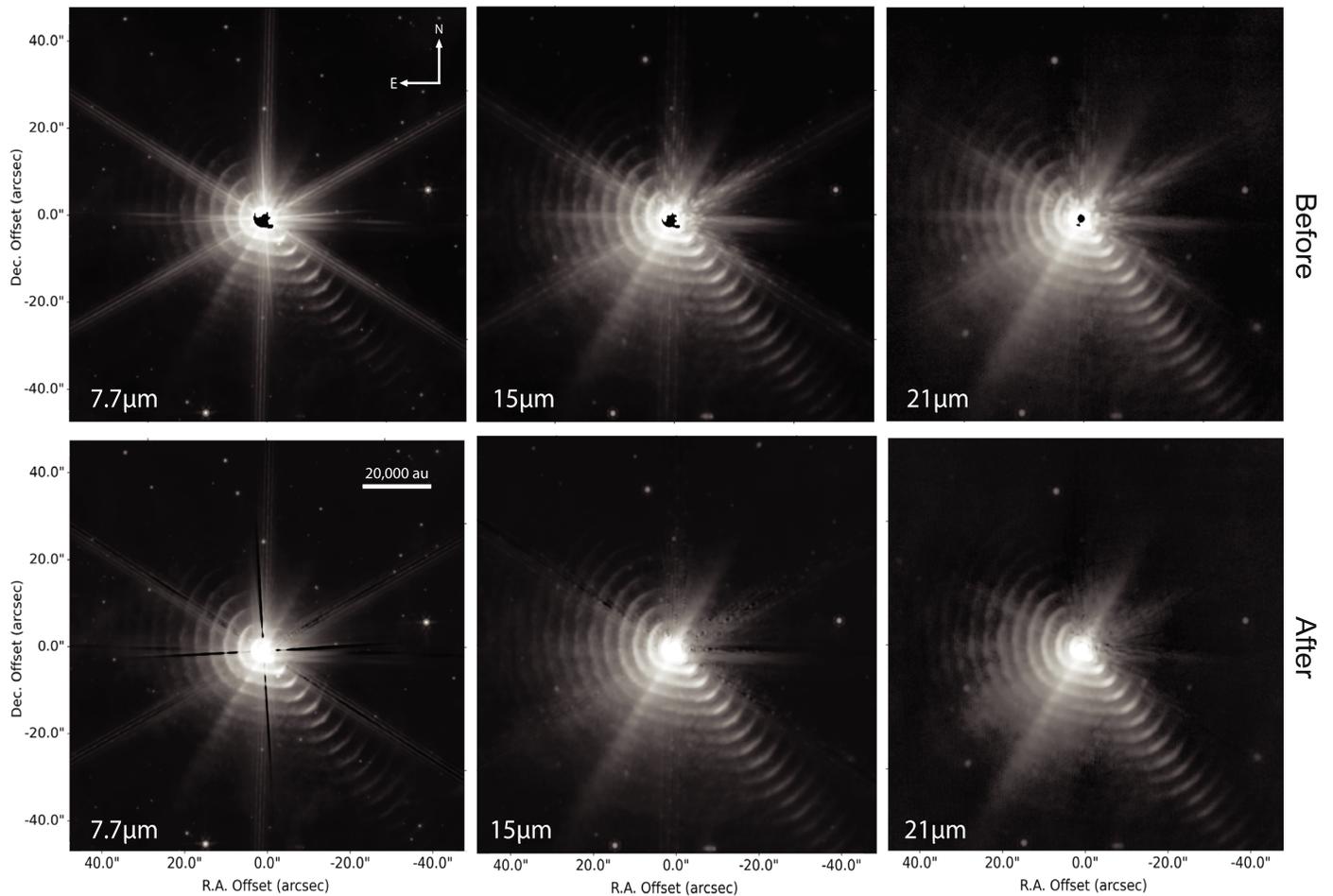

**Figure 6.** PSF subtraction. Before (top) and after (bottom) PSF subtraction of Cycle 1 images at 7.7, 15, and 21 μm. The varying flux of the diffraction spikes as a function of radius in the 7.7 μm image required a complex scaling of the PSF model which resulted in some areas of oversubtraction.

---

[25] https://github.com/emmalieb/jwstmiri_psfsubtraction

**Table 1**
Radial Flux Profiles—Displacements, Proper Motions, and Projected Velocities

| C1 Shell | $\Delta r_{7.7}$ (mas) | $PM_{7.7}$ (mas yr$^{-1}$) | $v_{7.7}$ (km s$^{-1}$) | $\Delta r_{15}$ (mas) | $PM_{15}$ (mas yr$^{-1}$) | $v_{15}$ (km s$^{-1}$) | $\Delta r_{21}$ (mas) | $PM_{21}$ (mas yr$^{-1}$) | $v_{21}$ (km s$^{-1}$) |
|---|---|---|---|---|---|---|---|---|---|
| 1 | 373 ± 20 | 334 ± 18 | 2595 ± 136 | 308 ± 10 | 275 ± 9 | 2139 ± 72 | 342 ± 8 | 306 ± 7 | 2377 ± 58 |
| 2 | 365 ± 24 | 327 ± 21 | 2540 ± 167 | 334 ± 9 | 298 ± 8 | 2320 ± 63 | 363 ± 5 | 325 ± 5 | 2526 ± 38 |
| 3 | 364 ± 29 | 325 ± 26 | 2528 ± 203 | 340 ± 9 | 304 ± 8 | 2365 ± 65 | 372 ± 6 | 332 ± 5 | 2584 ± 42 |
| 4 | 368 ± 22 | 329 ± 19 | 2562 ± 150 | 338 ± 23 | 302 ± 21 | 2352 ± 162 | 368 ± 11 | 329 ± 9 | 2560 ± 74 |
| 5 | 370 ± 19 | 331 ± 17 | 2575 ± 130 | 340 ± 15 | 304 ± 13 | 2365 ± 102 | 372 ± 7 | 333 ± 6 | 2587 ± 47 |
| 6 | 353 ± 32 | 316 ± 28 | 2455 ± 220 | 337 ± 14 | 301 ± 13 | 2345 ± 101 | 365 ± 9 | 327 ± 8 | 2540 ± 61 |
| 7 | 371 ± 22 | 332 ± 20 | 2581 ± 152 | 338 ± 20 | 302 ± 18 | 2350 ± 137 | 359 ± 8 | 321 ± 7 | 2496 ± 53 |
| 8 | 399 ± 20 | 356 ± 18 | 2771 ± 141 | 332 ± 14 | 297 ± 13 | 2309 ± 100 | 366 ± 7 | 327 ± 6 | 2547 ± 48 |
| 9 | 346 ± 22 | 310 ± 19 | 2408 ± 151 | 337 ± 22 | 301 ± 20 | 2345 ± 155 | 369 ± 19 | 330 ± 17 | 2566 ± 129 |
| 10 | 373 ± 40 | 333 ± 35 | 2590 ± 276 | 347 ± 20 | 310 ± 17 | 2412 ± 136 | 370 ± 9 | 331 ± 8 | 2572 ± 60 |
| 11 | 391 ± 21 | 349 ± 19 | 2718 ± 147 | 338 ± 9 | 302 ± 8 | 2349 ± 60 | 343 ± 19 | 307 ± 17 | 2385 ± 132 |
| 12 | 373 ± 19 | 333 ± 17 | 2594 ± 130 | 327 ± 12 | 293 ± 11 | 2276 ± 82 | 392 ± 12 | 351 ± 11 | 2729 ± 86 |
| 13 | 543 ± 24 | 485 ± 21 | 3773 ± 166 | 309 ± 19 | 276 ± 17 | 2148 ± 129 | 341 ± 24 | 305 ± 21 | 2369 ± 165 |
| 14 | 369 ± 17 | 330 ± 16 | 2568 ± 121 | 358 ± 19 | 320 ± 17 | 2488 ± 131 | 384 ± 47 | 343 ± 42 | 2672 ± 327 |
| 15 | 484 ± 30 | 433 ± 27 | 3365 ± 210 | 340 ± 17 | 304 ± 15 | 2363 ± 116 | 277 ± 29 | 248 ± 26 | 1928 ± 202 |
| 16 | 484 ± 54 | 432 ± 48 | 3363 ± 373 | 355 ± 19 | 317 ± 17 | 2466 ± 136 | | | |
| **E Shell** | $\Delta r_{7.7}$ (mas) | $PM_{7.7}$ (mas yr$^{-1}$) | $v_{7.7}$ (km s$^{-1}$) | $\Delta r_{15}$ (mas) | $PM_{15}$ (mas yr$^{-1}$) | $v_{15}$ (km s$^{-1}$) | $\Delta r_{21}$ (mas) | $PM_{21}$ (mas yr$^{-1}$) | $v_{21}$ (km s$^{-1}$) |
| 1 | 306 ± 20 | 274 ± 18 | 2131 ± 142 | 868 ± 56 | 776 ± 50 | [a] | 302 ± 20 | 270 ± 18 | 2102 ± 137 |
| 2 | 311 ± 28 | 278 ± 25 | 2164 ± 194 | 331 ± 30 | 296 ± 27 | 2304 ± 208 | 317 ± 32 | 283 ± 28 | 2203 ± 220 |
| 3 | 352 ± 37 | 315 ± 33 | 2449 ± 260 | 333 ± 19 | 298 ± 17 | 2319 ± 130 | 317 ± 21 | 284 ± 19 | 2206 ± 147 |
| 4 | 341 ± 44 | 305 ± 40 | 2372 ± 308 | 339 ± 40 | 303 ± 35 | 2358 ± 275 | 260 ± 39 | 233 ± 35 | 1811 ± 269 |
| 5 | 324 ± 92 | 289 ± 82 | 2252 ± 641 | 307 ± 34 | 275 ± 30 | 2136 ± 235 | 304 ± 38 | 271 ± 34 | 2111 ± 263 |
| 6 | 169 ± 87 | 151 ± 77 | 1176 ± 602 | 376 ± 18 | 336 ± 16 | 2616 ± 122 | 141 ± 84 | 126 ± 75 | 979 ± 581 |
| 7[a] | | | | | | | | | |
| 8 | 334 ± 70 | 298 ± 63 | 2322 ± 490 | 338 ± 93 | 302 ± 84 | 2347 ± 650 | | | |
| 9[a] | | | | | | | | | |
| 10 | 303 ± 73 | 271 ± 66 | 2108 ± 510 | 900 ± 103 | 804 ± 92 | [a] | | | |
| 11 | 371 ± 107 | 332 ± 96 | 2582 ± 745 | 265 ± 98 | 237 ± 87 | 1846 ± 679 | | | |
| **C0 Shell** | $\Delta r_{7.7}$ (mas) | $PM_{7.7}$ (mas yr$^{-1}$) | $v_{7.7}$ (km s$^{-1}$) | $\Delta r_{15}$ (mas) | $PM_{15}$ (mas yr$^{-1}$) | $v_{15}$ (km s$^{-1}$) | $\Delta r_{21}$ (mas) | $PM_{21}$ (mas yr$^{-1}$) | $v_{21}$ (km s$^{-1}$) |
| 1 | 288 ± 39 | 258 ± 35 | 2005 ± 271 | 205 ± 25 | 183 ± 22 | 1424 ± 173 | 271 ± 49 | 242 ± 44 | 1885 ± 340 |
| 2 | 293 ± 31 | 262 ± 28 | 2034 ± 219 | 267 ± 34 | 239 ± 31 | 1856 ± 237 | 254 ± 41 | 227 ± 36 | 1764 ± 282 |
| 3 | 292 ± 48 | 261 ± 43 | 2034 ± 334 | 252 ± 37 | 225 ± 33 | 1749 ± 259 | 256 ± 50 | 228 ± 45 | 1777 ± 350 |
| 4 | 265 ± 76 | 237 ± 68 | 1845 ± 527 | 285 ± 38 | 255 ± 34 | 1985 ± 261 | 266 ± 50 | 237 ± 45 | 1847 ± 350 |
| 5 | 281 ± 18 | 251 ± 16 | 1955 ± 124 | 264 ± 149 | 236 ± 133 | 1834 ± 1033 | 436 ± 175 | 390 ± 157 | 3031 ± 1218 |
| 6 | 292 ± 50 | 261 ± 45 | 2027 ± 351 | 269 ± 23 | 240 ± 20 | 1868 ± 157 | 263 ± 39 | 235 ± 35 | 1826 ± 270 |
| 7 | 237 ± 85 | 212 ± 76 | 1647 ± 589 | 334 ± 53 | 299 ± 47 | 2323 ± 367 | 201 ± 45 | 179 ± 40 | 1395 ± 311 |

[a] We omit here an outlier that is due to poor signal to noise in the E feature.
**Notes.** Shell numbers are the same as those described by R. M. Lau et al. (2022). $\Delta r_{7.7}$, $\Delta r_{15}$, and $\Delta r_{21}$ are the differences of each shell position from Cycle 1 to Cycle 2 in mas for 7.7, 15, and 21 μm, respectively. $PM_{7.7}$, $PM_{15}$, and $PM_{21}$ are the proper motions in mas yr$^{-1}$ for 7.7, 15, and 21 μm, respectively. $v_{7.7}$, $v_{15}$, and $v_{21}$ are the projected velocities at a distance of 1.64 kpc for for 7.7, 15, and 21 μm, respectively. Note that the E values have been deprojected.

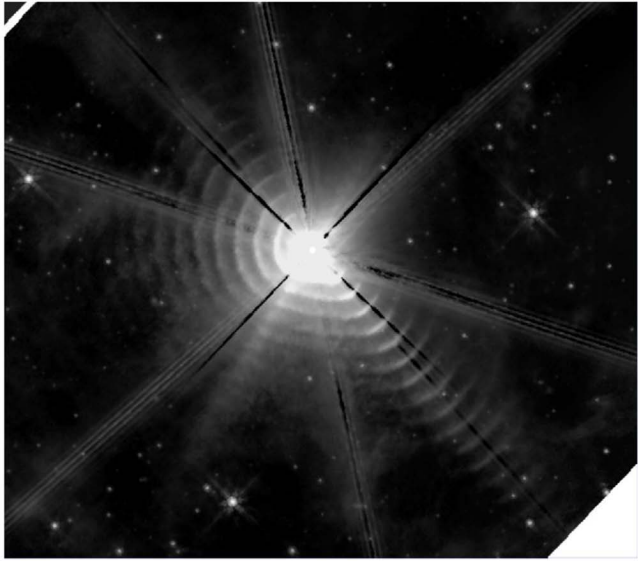

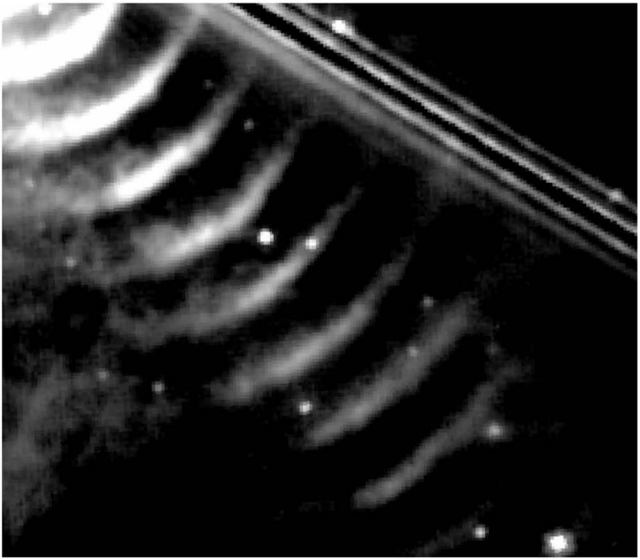

**Figure 7.** Animated videos. Top: video that blinks between the Cycle 1 and Cycle 2 images at 7.7 μm showing the entire system. Bottom: same as the top, but zoomed in on the C1 dust feature. All images have been PSF subtracted and the two cycles aligned.
(An animation of this figure is available in the online article.)

## ORCID iDs

Emma P. Lieb ⓘ https://orcid.org/0000-0002-4660-7452
Ryan M. Lau ⓘ https://orcid.org/0000-0003-0778-0321
Jennifer L. Hoffman ⓘ https://orcid.org/0000-0003-1495-2275
Macarena Garcia Marin ⓘ https://orcid.org/0000-0003-4801-0489
Takashi Onaka ⓘ https://orcid.org/0000-0002-8234-6747
Noel Richardson ⓘ https://orcid.org/0000-0002-2806-9339
Christopher M. P. Russell ⓘ https://orcid.org/0000-0002-9213-0763
Peredur M. Williams ⓘ https://orcid.org/0000-0002-8092-980X

## References

Astropy Collaboration, Price-Whelan, A. M., Lim, P. L., et al. 2022, ApJ, 935, 167
Bailer-Jones, C. A. L., Rybizki, J., Fouesneau, M., Mantelet, G., & Andrae, R. 2018, AJ, 156, 58
Bradley, L., Sipőcz, B., Robitaille, T., et al. 2024, astropy/photutils: v1.12.0, Zenodo, doi:10.5281/zenodo.10967176
Dicken, D., Marín, M. G., Shivaei, I., et al. 2024, A&A, 689, A5
Draine, B. T. 2011, Physics of the Interstellar and Intergalactic Medium (Princeton, NJ: Princeton Univ. Press)
Earl, N., Tollerud, E., O'Steen, R., et al. 2024, astropy/specutils: v1.18.0, Zenodo, doi:10.5281/zenodo.13942238
Eatson, J. W., Pittard, J. M., & Van Loo, S. 2022, MNRAS, 517, 4705
Eenens, P. R. J., & Williams, P. M. 1994, MNRAS, 269, 1082
Fahed, R., Moffat, A. F. J., Zorec, J., et al. 2011, MNRAS, 418, 2
Gaia Collaboration, Vallenari, A., Brown, A. G. A., et al. 2023, A&A, 674, A1
Gáspár, A., Rieke, G. H., Guillard, P., et al. 2020, PASP, 133, 014504
Han, Y., Tuthill, P. G., Lau, R. M., & Soulain, A. 2022, Natur, 610, 269272
Lau, R. M., Hankins, M. J., Han, Y., et al. 2022, NatAs, 6, 1308
Lau, R. M., Wang, J., Hankins, M. J., et al. 2023, ApJ, 951, 89
Lépine, S., Eversberg, T., & Moffat, A. F. J. 1999, AJ, 117, 1441
Lieb, E. 2024, JWST+MIRI Image PSF Diffraction Spike Subtraction, Zenodo, doi:10.5281/zenodo.14207684
Marchenko, S. V., & Moffat, A. F. J. 2007, in ASP Conf. Ser. 367, Massive Stars in Interactive Binaries, ed. N. St.-Louis & A. F. J. Moffat (San Francisco, CA: ASP), 213
Marchenko, S. V., Moffat, A. F. J., Ballereau, D., et al. 2003, ApJ, 596, 1295
Monnier, J. D., Tuthill, P. G., & Danchi, W. C. 2002, ApJL, 567, L137
Perrin, M. D., Soummer, R., Elliott, E. M., Lallo, M. D., & Sivaramakrishnan, A. 2012, Proc. SPIE, 8442, 84423D
Pittard, J. 2006, Mass-Loaded Flows, 245 (Dordrecht: Springer)
Pittard, J. M. 2007, ApJL, 660, L141
Rate, G., & Crowther, P. A. 2020, MNRAS, 493, 1512
Rest, A., Pierel, J., Correnti, M., et al. 2023a, arminrest/jhat: The JWST HST Alignment Tool (JHAT), v2, Zenodo, doi:10.5281/zenodo.7892935
Rest, A., Roberts-Pierel, J., Correnti, M., et al. 2023b, AAS Meeting Abstracts, 241, 358.02
Thomas, J. D., Richardson, N. D., Eldridge, J. J., et al. 2021, MNRAS, 504, 5221
Tuthill, P. G., Monnier, J. D., & Danchi, W. C. 1999, Natur, 398, 487
Usov, V. V. 1991, MNRAS, 252, 49
Williams, P. 2011, arXiv:1101.1046
Williams, P. M., & Eenens, P. R. J. 1989, MNRAS, 240, 445
Williams, P. M., Marchenko, S. V., Marston, A. P., et al. 2009, MNRAS, 395, 1749
Wright, G. S., Wright, D., Goodson, G. B., et al. 2015, PASP, 127, 595